\documentclass[9pt,twocolumn]{article}
\usepackage[margin=0.7in]{geometry}
\usepackage[utf8]{inputenc}
\usepackage{textcomp}
\usepackage{microtype}
\usepackage{graphicx}
\usepackage{subcaption}
\usepackage{booktabs}
\usepackage{colortbl}
\usepackage{xcolor}
\usepackage{multirow}
\usepackage{adjustbox}
\usepackage{nicefrac}

\usepackage{amsmath}
\usepackage{amssymb}
\usepackage{amsfonts}
\usepackage{mathtools}
\usepackage{amsthm}

\usepackage{algorithm}
\usepackage{algorithmic}
\usepackage{authblk}
\usepackage[numbers]{natbib}
\usepackage{hyperref}
\usepackage{url}
\usepackage[capitalize,noabbrev]{cleveref}

\definecolor{ourscolor}{gray}{0.92}

\usepackage{float}
\usepackage{stfloats}

\theoremstyle{plain}

\theoremstyle{definition}

\theoremstyle{remark}

\usepackage{mathpazo}  % Palatino font for text and math
\usepackage[T1]{fontenc}

\newcommand{\lp}[1]{\textcolor{brown}{[LP: #1]}}

\title{S-JEPA : Soft Clustering Anchors for Self-Supervised Speech Representation Learning}

\author{
  \textbf{Georgios Ioannides}$^{1,3,7}$ \quad
  \textbf{Adrian Kieback}$^{3,*}$ \quad
  \textbf{Judah Goldfeder}$^{4,*}$ \quad
  \textbf{Linsey Pang}$^{5}$\\
  \textbf{Aman Chadha}$^{3,6}$ \quad
  \textbf{Aaron Elkins}$^{3}$ \quad
  \textbf{Yann LeCun}$^{2}$ \quad
  \textbf{Ravid Shwartz-Ziv}$^{2}$\vspace{1em}\\
  {\small
  $^{1}$Carnegie Mellon University \quad
  $^{2}$New York University \quad
  $^{3}$James Silberrad Brown Center for AI\\
  $^{4}$Columbia University \quad
  $^{5}$Northeastern University \quad
  $^{6}$Stanford University \quad
  $^{7}$Amazon GenAI$^{\dagger}$ \vspace{0.5em}\\
  $^{*}$Equal contribution \quad Correspondence: \texttt{gioannid@alumni.cmu.edu}
  }
}

\date{}

\begin{document}

\maketitle
\footnotetext[7]{$^{\dagger}$Work does not relate to position at Amazon.}
\begin{abstract}

Self-supervised speech encoders are predominantly trained by predicting discrete hard cluster IDs at masked positions, a recipe that collapses acoustic ambiguity at category boundaries and requires interrupting training to re-cluster the entire corpus between iterations. We introduce \textbf{S-JEPA}, a JEPA-style encoder-predictor pair trained to match the soft posteriors of a Gaussian Mixture Model at masked positions via KL divergence. Training runs as one continuous optimization trajectory in two phases: a fixed GMM over MFCC features, then an \emph{online} GMM over encoder features, with the input layer selected adaptively from a label-free signal, removing both the offline re-cluster step and the hand-tuned choice of which transformer layer to cluster on. Under the SUPERB protocol, S-JEPA achieves the lowest WER among evaluated SSL methods below 90M parameters and matches HuBERT-Base on emotion recognition at roughly half its parameter count, establishing a new Pareto frontier without offline re-clustering or teacher distillation. An analysis of the predictor's per-frame entropy on held-out speech reveals a bimodal distribution with a substantial minority of frames near the entropy of a perfect two-cluster tie, providing direct empirical evidence that the soft-target objective preserves the acoustic ambiguity that hard targets would collapse. Code is available at \url{https://github.com/gioannides/s-jepa}.

\end{abstract}

%==============================================================================
\section{Introduction}
%==============================================================================

Self-supervised speech representation learning is currently dominated by a single recipe: cluster acoustic features offline, then train an encoder to predict cluster IDs at masked positions via cross-entropy. HuBERT~\citep{hsu2021hubert} introduced this pattern with two iterations of $k$-means (first on MFCC features, then on intermediate transformer features), and WavLM~\citep{chen2022wavlm} extended it with denoising augmentation. The recipe works, but it has two awkward properties. First, hard cluster assignment collapses the acoustic ambiguity at category boundaries~\citep{ladefoged2014course,stevens1998acoustic} into a single ID. Frames at phone boundaries, transitions, and silence/non-silence shifts are forced into one partition. Second, the recipe requires a stop-restart training pipeline: training is interrupted between iterations to re-fit cluster centers over the entire corpus.

We introduce \textbf{S-JEPA}, a self-supervised speech objective that addresses both points within a single training pass. S-JEPA trains a JEPA-style encoder--predictor pair~\citep{lecun2022jepa,assran2023ijepa} to match the soft posteriors of a Gaussian Mixture Model (GMM)~\citep{dempster1977em,bishop2006pattern} at masked positions, supervised by a single KL divergence loss. Training proceeds in two phases mirroring HuBERT's clustering iterations: a GMM over MFCC features, then a GMM over encoder features. Phase~2 fits its GMM \emph{online} from the EMA encoder's outputs and selects the GMM's input layer adaptively by tracking effective rank, eliminating the offline re-cluster step. With a 51.8M-parameter encoder, S-JEPA reaches \textbf{12.10\% WER on LibriSpeech test-clean} under the SUPERB ASR protocol (greedy CTC decoding; 8.50\% with 4-gram LM rescoring), the lowest of any evaluated SSL method below 90M parameters, and \textbf{64.83\% accuracy on emotion recognition}, approximately matching HuBERT-Base at 55\% of its parameter count. The online updates fold the GMM's cost into the existing training forward pass, removing the full-corpus relabel pass that HuBERT and WavLM use between iterations.

The soft-target objective is also analytically informative. On 153K held-out frames, the predictor's per-frame entropy is bimodal: a confident regime concentrated below 0.3~bits, and a clear secondary mode near 1~bit corresponding to two-way ties between competing clusters. Over a third of frames carry more than 1~bit of entropy. These are frames where a hard-target objective would have to commit to one cluster ID and discard the inter-cluster mass. This is direct empirical evidence that the soft-target objective preserves uncertainty at acoustic boundaries rather than collapsing it, and it characterizes \emph{what} the encoder learned without us having to claim it explains \emph{why} the method is parameter-efficient.

\begin{figure*}[!t]
\centering
\includegraphics[width=0.7\linewidth,height=0.3\textheight]{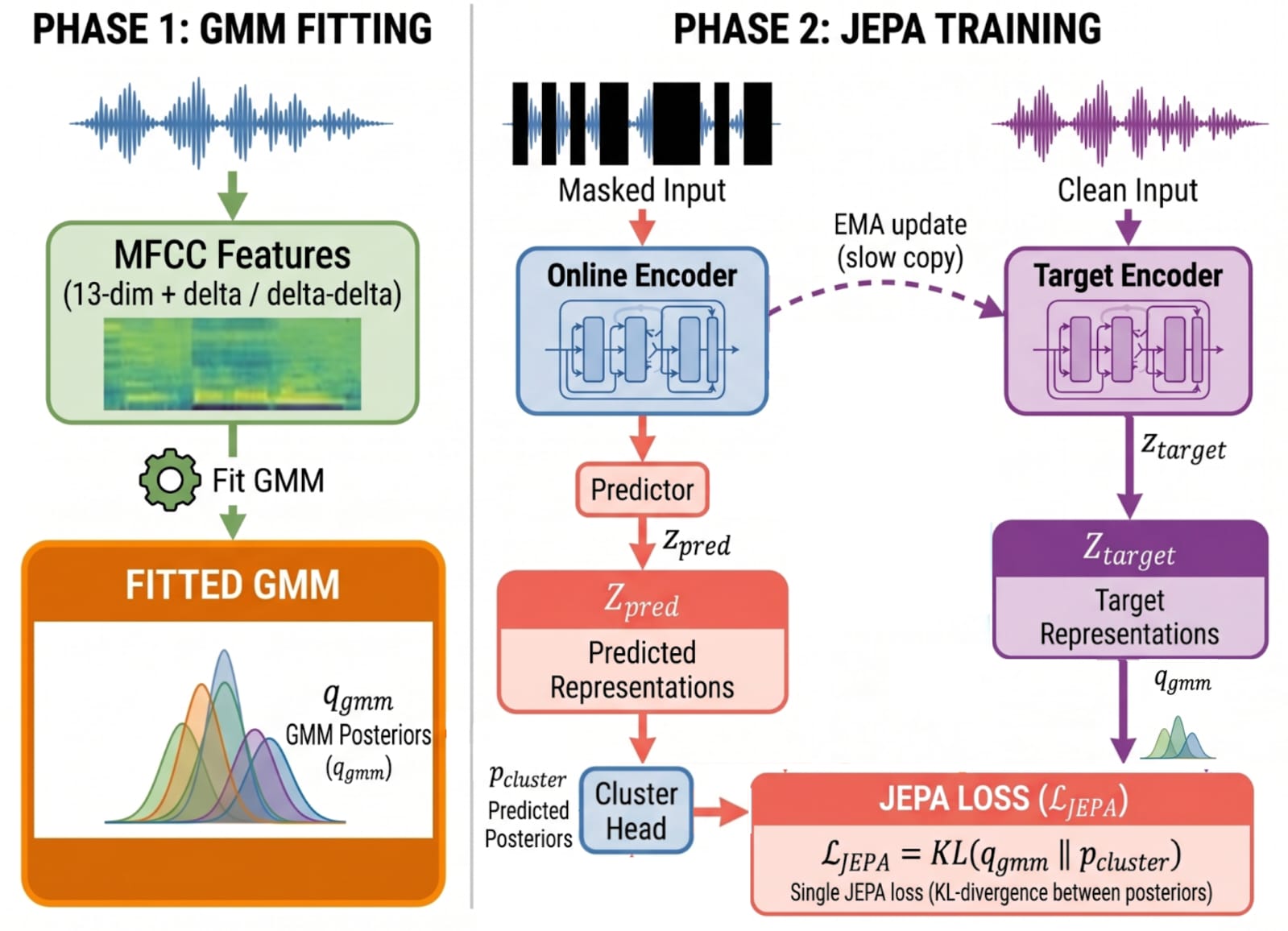}
\caption{\textbf{S-JEPA: a JEPA-style encoder--predictor pair matches soft GMM posteriors at masked positions, with the GMM fit in Phase~1 (frozen, over MFCC features) and updated online in Phase~2 (over EMA encoder features).} The single training signal is KL divergence at masked positions; the predictor, cluster head, and GMM are discarded after pre-training.}
\label{fig:overview}
\end{figure*}

\paragraph{Contributions.}
\begin{itemize}
\item \textbf{A soft-target masked-prediction objective for speech SSL.} 
We replace HuBERT's hard $k$-means labels and cross-entropy loss with the 
soft posteriors of a GMM over acoustic features, matched via KL 
divergence at masked positions. The objective uses a single-loss and is
architecturally compatible with the JEPA pattern (cf. Section~\ref{sec:method}).

\item \textbf{A single-pass training pipeline that eliminates offline 
re-clustering.} Phase 2 fits its GMM online from minibatch sufficient 
statistics, selects the GMM's input layer adaptively by effective rank, 
and uses a target-encoder EMA decay that alternates between two values 
rather than holding fixed. The alternating EMA serves as a 
continuous-time analogue of HuBERT's offline re-cluster step. The two 
phases run as one continuous optimization trajectory 
(cf. Section~\ref{sec:two_phase}).

\item \textbf{A new Pareto frontier on SUPERB below 90M parameters, with 
structured boundary uncertainty.} S-JEPA is trained directly at 51.8M 
parameters in a single pass, without a pretrained teacher and without 
offline re-clustering, and on three SUPERB tasks we evaluate it dominates 
all SSL methods below 90M while matching HuBERT-Base on emotion 
recognition at 55\% of its parameter count (cf. Section~\ref{sec:experiments}). 
The predictor's per-frame entropy on held-out speech is bimodal, with a 
substantial minority of frames in a structured two-way-tie regime that a 
hard-target objective could not represent (cf. Section~\ref{sec:analysis}).
\end{itemize}

\section{Related Work}
\label{sec:related}

%==============================================================================

\paragraph{Speech SSL families.} Self-supervised speech encoders fall into two broad families. \emph{Contrastive} methods predict future or masked latents and contrast against negatives: CPC~\citep{oord2018cpc}, wav2vec~\citep{schneider2019wav2vec}, and wav2vec~2.0~\citep{baevski2020wav2vec}, which quantizes the latent space. \emph{Masked-prediction} methods reframe SSL as classification over offline cluster IDs at masked positions: HuBERT~\citep{hsu2021hubert} clusters MFCC features and then intermediate transformer features via offline $k$-means and supervises with cross-entropy; WavLM~\citep{chen2022wavlm} adds denoising augmentation and a gated relative position bias; w2v-BERT~\citep{chung2021w2vbert} combines contrastive and masked-prediction losses. A separate line of work avoids explicit target discretization with EMA teachers~\citep{grill2020byol,caron2021dino,baevski2022data2vec}, supervising via regression on continuous teacher representations. JEPAs~\citep{lecun2022jepa,assran2023ijepa,fang2024ajepa} predict representations across views in a learned latent space. S-JEPA follows the JEPA architectural pattern (encoder, mask token, predictor) and the masked-prediction framing, but supervises with soft categorical targets rather than hard cluster IDs or continuous regression targets.

\paragraph{Clustering methods.} Two methods sit closest to S-JEPA and clarify what it does and does not do. HuBERT shares S-JEPA's discrete categorical target structure and the two-iteration recipe at the cardinality level, but assigns each frame to a single cluster (hard $k$-means) and refits cluster centers offline between iterations; S-JEPA replaces both choices, using soft GMM posteriors on the target side and online GMM updates on the optimization side. BEST-RQ~\citep{chiu2022bestrq} addresses the same offline re-cluster cost with a different solution: it freezes a random-projection quantizer for the entire training run, removing the iterative re-cluster step by removing the iteration. S-JEPA keeps the iterative structure but lets it run continuously, with the GMM updated online from the encoder's own evolving features. Closest of all is our own prior work~\citep{ioannides2026softanchors}, which anchors a JEPA speech encoder with a \emph{frozen} GMM fit once on log-mel features, consumed as an auxiliary KL target under a decaying weight alongside a latent JEPA regression loss. S-JEPA differs in three respects: the GMM is updated \emph{online} from evolving encoder features rather than frozen on log-mels; the soft posteriors are the \emph{single} training target, with no separate latent-regression term and no $\lambda$ schedule to tune; and the GMM's input layer is selected adaptively rather than fixed.

\paragraph{Distillation methods.} A separate line of work targets smaller speech encoders by distilling a 
pretrained model: DistilHuBERT~\citep{chang2022distilhubert}, FitHuBERT, 
and LightHuBERT all compress a $\sim$95M-parameter HuBERT teacher into a 
smaller student. The student's parameter count is the inference-time 
cost of the compressed model, not the training-time cost of producing 
it: each of these methods inherits the full HuBERT pretraining pipeline 
as a prerequisite. S-JEPA reaches the same parameter regime through 
pretraining alone, with no teacher.

\paragraph{Lineage.} Soft cluster supervision via Gaussian mixtures~\citep{dempster1977em,bishop2006pattern} matched by KL divergence~\citep{tishby2000information} is classical. We are not proposing a new clustering method, prediction architecture, or speech encoder backbone; the contribution is the recipe that combines soft GMM posteriors, online updates, and JEPA-style masked prediction into a single-pass speech SSL training pipeline.
 
%==============================================================================
\section{Method}
%==============================================================================
\label{sec:method}
\subsection{Architecture}
\label{sec:arch}

The architecture follows the JEPA pattern with three trained components --- an encoder $f_\phi$, a predictor $h_\psi$, and a cluster head $g_\omega$ --- plus a non-trained auxiliary component used only in Phase~2: an \emph{EMA encoder} $\bar{f}$, a frozen architectural copy of $f_\phi$ whose parameters are an exponential moving average of those of $f_\phi$. The EMA encoder supplies the input features to the online GMM in Phase~2 (cf. Section~\ref{sec:two_phase}); it is unused in Phase~1. The encoder $f_\phi$ maps the raw waveform to frame-level representations. The predictor takes the encoder output and produces frame-level predictions, with a learned mask token injected at masked positions and learned positional embeddings added before its Transformer layers. The cluster head is a shared MLP that maps frame-level representations to $K$-dimensional logits and is applied both to the encoder output (at visible positions) and to the predictor output (at masked positions); only the latter contributes to the training loss.

Concretely, let $\text{mask} \in \{0,1\}^T$ be a binary mask marking visible (1) and masked (0) positions. Let $z = f_\phi(x)$ be the encoder output and let $\tilde{z}$ be $z$ with masked positions zeroed out. The predictor injects a learned mask token at the masked positions of $\tilde{z}$, adds positional embeddings, and runs a small Transformer (a single 768-dimensional encoder layer with 8 attention heads) to produce $\hat{z} = h_\psi(\tilde{z}, \text{mask})$. Cluster logits are then computed at every position by $g_\omega$:
\begin{align}
u^{\text{vis}}_t &= g_\omega(z_t) \;\; \text{\small(visible positions; not supervised)} \\
u^{\text{mask}}_t &= g_\omega(\hat{z}_t) \;\; \text{\small(masked positions; supervised in Equation~\ref{eq:loss})}
\end{align}
We write $p_t = \text{softmax}(u^{\text{mask}}_t)$ for the predictor's softmaxed output at masked frame $t$, which is what is matched against the GMM posterior in the training loss. After training, only $f_\phi$ is retained for downstream use; the predictor, cluster head, and GMM are discarded. Full architectural details are in Section~\ref{app:arch} of the Appendix.

\iffalse 
Concretely, let $\mathit{mask} \in \{0,1\}^T$ be a binary mask marking visible (1) and masked (0) positions. Let $z = f_\phi(\mathbf{x})$ be the encoder output, computed with $z$ zeroed out at masked positions. The predictor injects a learned mask token at masked positions in $z$, adds positional embeddings, and runs a small Transformer to produce $\hat{z} = h_\psi(z, \mathit{mask})$. Cluster logits are then computed at every position by $g_\omega$:
\begin{align}
\mathbf{u}^{\text{vis}}_t  &= g_\omega(z_t)        & \text{(visible positions; not supervised)} \\
\mathbf{u}^{\text{mask}}_t &= g_\omega(\hat{z}_t)  & \text{(masked positions; supervised in \Cref{eq:loss})}
\end{align}
We write $\mathbf{p}_t = \mathrm{softmax}(\mathbf{u}^{\text{mask}}_t)$ for the predictor's softmaxed output at masked frame $t$, which is what is matched against the GMM posterior in the training loss. After training, only $f_\phi$ is retained for downstream use; the predictor, cluster head, and GMM are discarded. Full architectural details are in \Cref{app:architecture}.
\fi

\subsection{Soft Targets and KL Loss}
\label{sec:targets_loss}

A $K$-component diagonal-covariance GMM~\citep{dempster1977em,bishop2006pattern} provides soft posteriors over an acoustic feature space:
\begin{equation}
q_k(m) = \frac{\pi_k \mathcal{N}(m; \mu_k, \sigma^2_k)}{\sum_j \pi_j \mathcal{N}(m; \mu_j, \sigma^2_j)}.
\label{eq:gmm_posterior}
\end{equation}
At each frame $t$, an acoustic feature vector $m_t$ is computed under \texttt{no\_grad} (from 39-dimensional MFCC features in Phase~1, or from an EMA encoder in Phase~2; see Section~\ref{sec:two_phase}), and the corresponding posterior $q_t = (q_1(m_t), \ldots, q_K(m_t))$ is used as the soft target. These posteriors assign nonzero probability to multiple components, so frames near a category boundary are not forced into a single partition.

\iffalse
A $K$-component diagonal-covariance GMM~\citep{dempster1977em, bishop2006pattern} provides soft posteriors over an acoustic feature space:
\begin{equation}
q_k(\mathbf{m}) = \frac{\pi_k \mathcal{N}(\mathbf{m}; \boldsymbol{\mu}_k, \boldsymbol{\sigma}_k^2)}{\sum_{j} \pi_j \mathcal{N}(\mathbf{m}; \boldsymbol{\mu}_j, \boldsymbol{\sigma}_j^2)}.
\end{equation}
At each frame $t$, an acoustic feature vector $\mathbf{m}_t$ is computed under \texttt{no\_grad} (from 39-dimensional MFCC features in Phase 1, or from an EMA encoder in Phase 2; see \Cref{sec:two-phase}), and the corresponding posterior $\mathbf{q}_t = (q_1(\mathbf{m}_t), \dots, q_K(\mathbf{m}_t))$ is used as the soft target. These posteriors assign nonzero probability to multiple components, so frames near a category boundary are not forced into a single partition.
\fi

The training loss is the KL divergence between the predictor's softmaxed output and the GMM posterior, averaged over a subset of frame positions $S$:
\begin{equation}
\mathcal{L} = \frac{1}{|S|} \sum_{t \in S} \mathrm{KL}\!\left(q_t \,\|\, p_t\right),
\label{eq:loss}
\end{equation}
where both arguments are distributions over the same $K$ GMM components. The set $S$ is either the masked positions $\mathcal{M}$, or the union of masked and visible positions, depending on the training stage (cf. Section~\ref{sec:two_phase}).

\iffalse
The training loss is a KL divergence between the predictor's softmaxed output and the GMM posterior, computed on a subset of frame positions $\mathcal{S}$:
\begin{equation}
\mathcal{L} \;=\; \frac{1}{|\mathcal{S}|} \sum_{t \in \mathcal{S}} \text{KL}\bigl(\mathbf{q}_t \,\|\, \mathbf{p}_t\bigr),
\label{eq:loss}
\end{equation}
where the two arguments are distributions over the same $K$ GMM components. The set $\mathcal{S}$ is either the masked positions $\mathcal{M}$, or the union of masked and visible positions, depending on the training stage --- see \Cref{sec:two-phase}.
\fi

\subsection{Two-Phase Training}
\label{sec:two_phase}

Training proceeds in two phases that run as one continuous optimization trajectory.

\paragraph{Phase 1: MFCC GMM.} A $K{=}100$ diagonal-covariance GMM is fit once over 39-dimensional MFCC features (13 cepstral coefficients with first and second deltas) from the training corpus, initialized with mini-batch $k$-means~\citep{macqueen1967kmeans} on a reservoir-sampled subset and refined with expectation-maximization~\citep{dempster1977em}. The GMM is then held fixed throughout Phase~1. At each training step, GMM posteriors are computed on the MFCCs of each utterance and used as targets in Equation~\ref{eq:loss}. Phase~1 applies the loss at both masked and visible positions ($S = \mathcal{M} \cup \mathcal{V}$) with denoising augmentation enabled. Full details are provided in Section~\ref{app:gmm} of the Appendix.

\paragraph{Phase 2: online encoder-feature GMM.} At the Phase~1$\to$2 boundary, the EMA encoder $\bar{f}$ is initialized from the Phase~1 encoder, and a $K{=}500$ GMM is initialized over $\bar{f}$'s frame-level features at the active layer using the same mini-batch $k$-means + EM procedure as in Phase~1. Phase~2 begins with the same loss and augmentation configuration as Phase~1; partway through it transitions to a masked-only loss with augmentation off (cf. Appendix Section~\ref{app:phase2_transition}). Three departures from HuBERT's recipe in Phase~2 are described in the following paragraphs. $K{=}100$ and $K{=}500$ were selected to match the cluster cardinalities used in HuBERT iterations 1 and 2 respectively.

\paragraph{Online GMM updates.} Updating the GMM online lets it track the encoder's features as they evolve, rather than freezing a snapshot at a re-cluster boundary. After each minibatch we compute responsibility-weighted means and variances of the EMA encoder's frame-level features and EMA-update the GMM parameters with a slow decay; no gradient flows through the GMM. The per-step cost is the GMM E and M steps over the minibatch, which run alongside the existing training forward pass, so no separate corpus-scale re-label pass is required between iterations. Full update equations are provided in Section~\ref{app:gmm} of the Appendix.

\paragraph{Adaptive layer selection.} The encoder layer used as the GMM's input is selected adaptively rather than fixed by hand. Let $Z \in \mathbb{R}^{N \times D}$ denote a centered batch of $N{=}2{,}000$ frame embeddings of dimension $D$ from a given encoder layer, with singular values $\sigma_1 \geq \cdots \geq \sigma_D$ and normalized distribution $p_i = \sigma_i / \sum_j \sigma_j$. The \emph{effective rank} of $Z$ is
$\mathrm{erank}(Z) = \exp\!\left(-\sum_{i=1}^{D} p_i \log p_i\right)
\label{eq:erank}
$,
the exponential of the Shannon entropy of the normalized singular value spectrum. It is high when many singular values are comparable and low when dominated by a few large values; we use it as a label-free proxy for representational richness at each layer. Effective rank is a lower bound on matrix-based entropy in the Shannon case~\citep[Theorem~1]{skean2025layer}, and spread-of-spectrum measures of this family have been shown to correlate with downstream performance across transformer and state-space architectures, including their use as label-free layer-selection signals~\citep{skean2025layer,garrido2023rankme}. At a fixed cadence during Phase~2, we compute $\mathrm{erank}(Z^{(\ell)})$ at each transformer layer on a batch of EMA-encoder features, EMA-smooth across measurement intervals, and select the argmax as the GMM's input layer. When the argmax changes, the GMM continues updating online from the new layer with no offline re-cluster step. We additionally calibrated effective rank against per-layer downstream WER at the end of Phase~1 in our specific setting (cf. Appendix Section~\ref{app:erank_calibration}); we discuss the scope of that calibration in Section~\ref{sec:limitations}.
 
\begin{figure*}[!t]
\centering
\begin{subfigure}[t]{0.48\linewidth}
    \centering
    \includegraphics[width=\linewidth]{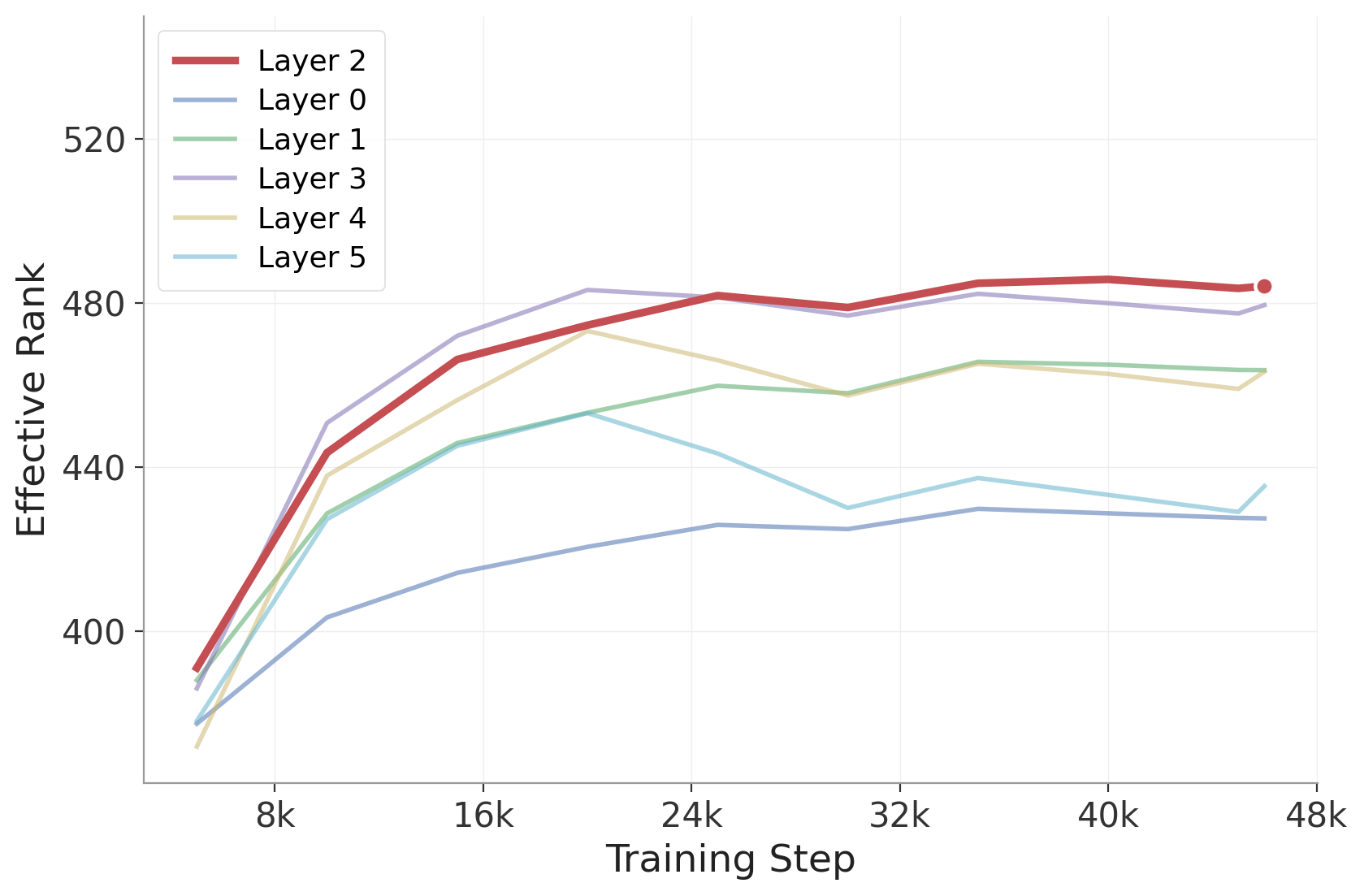}
    \caption{\textbf{Phase 1.} Layer 2 settles as the most expressive layer, motivating its use as the initial GMM input layer for Phase 2.}
    \label{fig:erank-phase1}
\end{subfigure}\hfill
\begin{subfigure}[t]{0.48\linewidth}
    \centering
    \includegraphics[width=\linewidth]{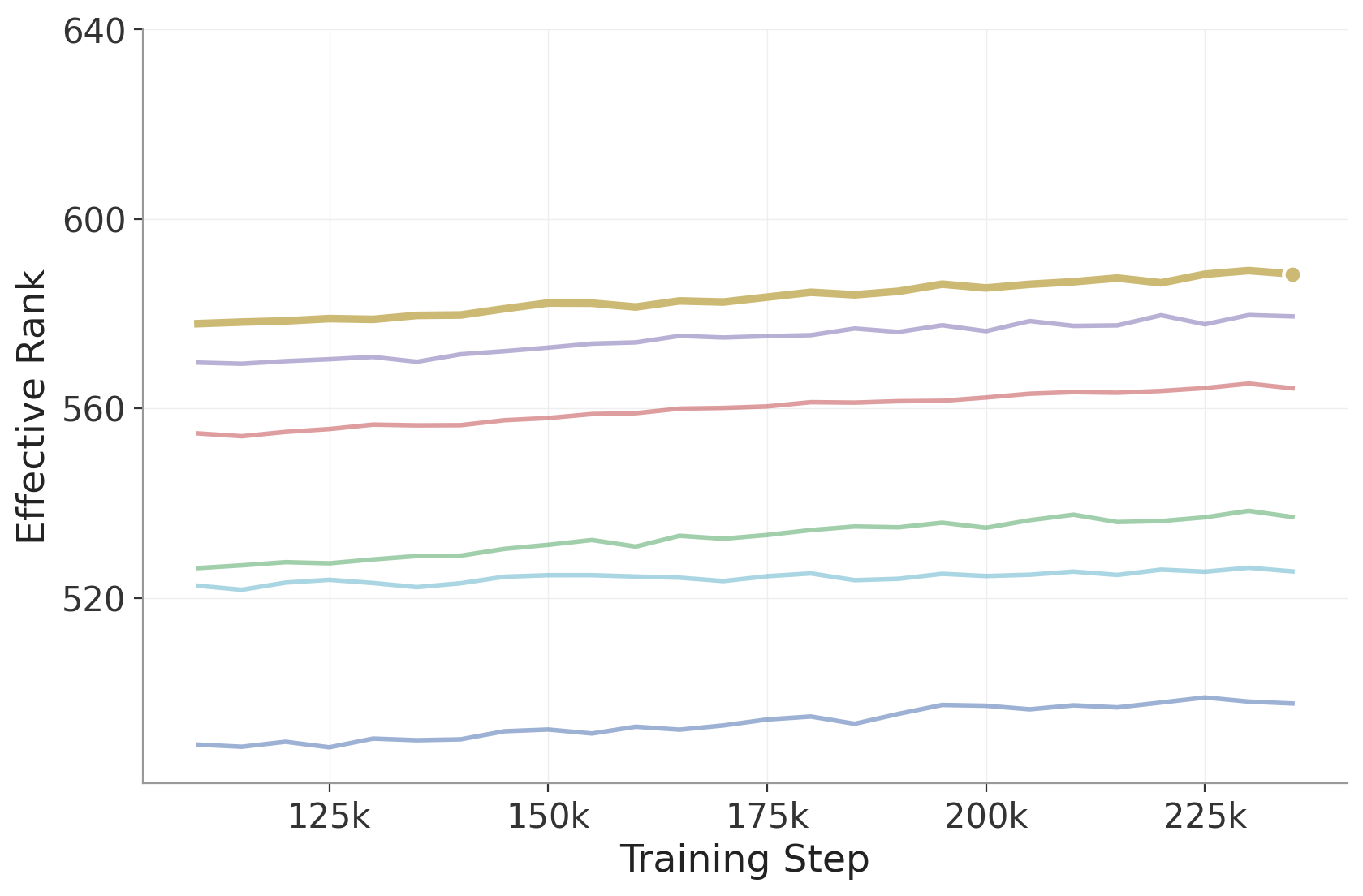}
    \caption{\textbf{Phase 2.} Under the online GMM, layer 4 dominates from the first measured checkpoint and the L4--L2 gap widens.}
    \label{fig:erank-phase2}
\end{subfigure}
\caption{Per-layer effective rank across both phases of training. The horizontal axis shows training step}
\label{fig:erank-phases}
\end{figure*}

In our 6-layer encoder (indexes 0-5), the active layer was initialized at $\ell^\star = 2$ (the highest-effective-rank layer index at the end of Phase~1) and the adaptive rule transitioned it to $\ell^\star = 4$ midway through Phase~2, where it remained (cf. Figure~\ref{fig:erank-phases}). Tracking development-set WER throughout Phase~2, we observed that once layer~2 ceased to be the highest-effective-rank layer, anchoring the GMM at $\ell^\star = 2$ was correlated with stalled WER improvement; switching to $\ell^\star = 4$ was correlated with continued progress. We do not claim higher effective rank causes better downstream performance, only that anchoring at the current argmax is correlated with continued progress.

\paragraph{Periodically switched EMA decay.} \label{par:cyclic-ema} A standard self-distillation setup uses a fixed slow EMA decay (e.g., $\alpha = 0.9999$) so the EMA encoder is nearly stationary~\citep{grill2020byol,baevski2022data2vec}. A stationary EMA encoder is the right choice when its outputs are continuous-valued teacher representations consumed directly as regression targets, but in our setup it interacts with the following issue: the EMA encoder's outputs are the input features to the online GMM, whose components are themselves estimated from those features. If the EMA encoder is held very slow throughout, the GMM never sees recent improvements in the online encoder; if it is held fast, the target distribution shifts on every step. We address this by switching the EMA decay rate periodically between two fixed values:
\begin{equation}
\alpha_t \in \{\alpha_{\text{fast}}, \alpha_{\text{slow}}\}, \quad \alpha_{\text{fast}} = 0.999, \quad \alpha_{\text{slow}} = 0.9999,
\label{eq:ema_switch}
\end{equation}
applied to the update $\bar{\phi} \leftarrow \alpha_t \bar{\phi} + (1 - \alpha_t)\phi$. Training begins at $\alpha_{\text{fast}}$; at a fixed step cadence the rate is flipped. Under $\alpha_{\text{fast}}$, the EMA encoder tracks the online encoder closely and the GMM's input distribution shifts to reflect recent encoder improvements; under $\alpha_{\text{slow}}$, that vocabulary is held approximately stable and the online encoder can learn against it without target drift swamping the gradient signal. Switching between two fixed rates separates these two requirements in time. Full schedule details are provided in Section~\ref{app:cyclic-ema}.

\iffalse
\paragraph{Holding a once-optimal anchor fixed is correlated with stalled progress.}
The shift in which layer carries the highest effective rank coincides with a change in downstream behaviour. We tracked WER on a held-out development set throughout Phase 2 and observed that, once layer 2 ceased to be the highest-effective-rank layer, anchoring the GMM at $\ell^\star\!=\!2$ produced visibly slower WER improvement than switching to the new argmax $\ell^\star\!=\!4$. After the adaptive rule transitioned the anchor to layer 4, downstream WER continued to decrease. We do not claim higher effective rank causes better downstream performance; the observation is that anchoring at a layer no longer at the argmax is correlated with stalled WER improvement, while anchoring at the current argmax is correlated with continued progress.

\paragraph{Calibration of effective rank as a layer-selection metric.}
At the end of Phase 1, we trained a CTC ASR head~\citep{graves2006ctc} on each of the 6 transformer layers (encoder frozen, head trained from scratch for 1{,}000 steps) and recorded WER on a held-out development set. The lowest-WER layer coincided with the highest-effective-rank layer on a fixed batch of features. We use this coincidence to motivate effective rank as our layer-selection signal during Phase 2; see \Cref{sec:limitations} for the scope of this calibration.
\fi
\subsection{Training Algorithm}
\label{sec:algorithm}

\Cref{alg:training} summarizes the full procedure. Block masking samples spans of frames until a target fraction is masked. The encoder $f_\phi$ and predictor $h_\psi$ are trained jointly via the gradient of Equation~\ref{eq:loss}. An EMA copy $\bar{f}$ is maintained throughout training; in Phase~2 its decay rate $\alpha_t$ is periodically switched between $\alpha_{\text{fast}}$ and $\alpha_{\text{slow}}$ as in Equation~\ref{eq:ema_switch}. In Phase~1 the GMM is the frozen MFCC GMM and the EMA encoder is unused.

\section{Experiments}

%==============================================================================
\label{sec:experiments}

\paragraph{Setup.} We pre-train S-JEPA on a large-scale English corpus combining LibriLight~\citep{kahn2020librilight} and an English subset of Granary~\citep{bai2024granary}, a total of approximately 83,000 hours, and evaluate it with the encoder frozen on three SUPERB benchmark suite tasks~\citep{yang2021superbspeechprocessinguniversal}. Following the SUPERB protocol, the encoder is frozen and small task-specific heads are trained on top. The encoder backbone uses 6 Transformer layers (51.8M parameters), targeting the sub-90M parameter regime; full hyperparameters are provided in Section~\ref{app:hyperparams} of the Appendix.

\begin{figure*}[!t]
\centering
\includegraphics[width=\linewidth]{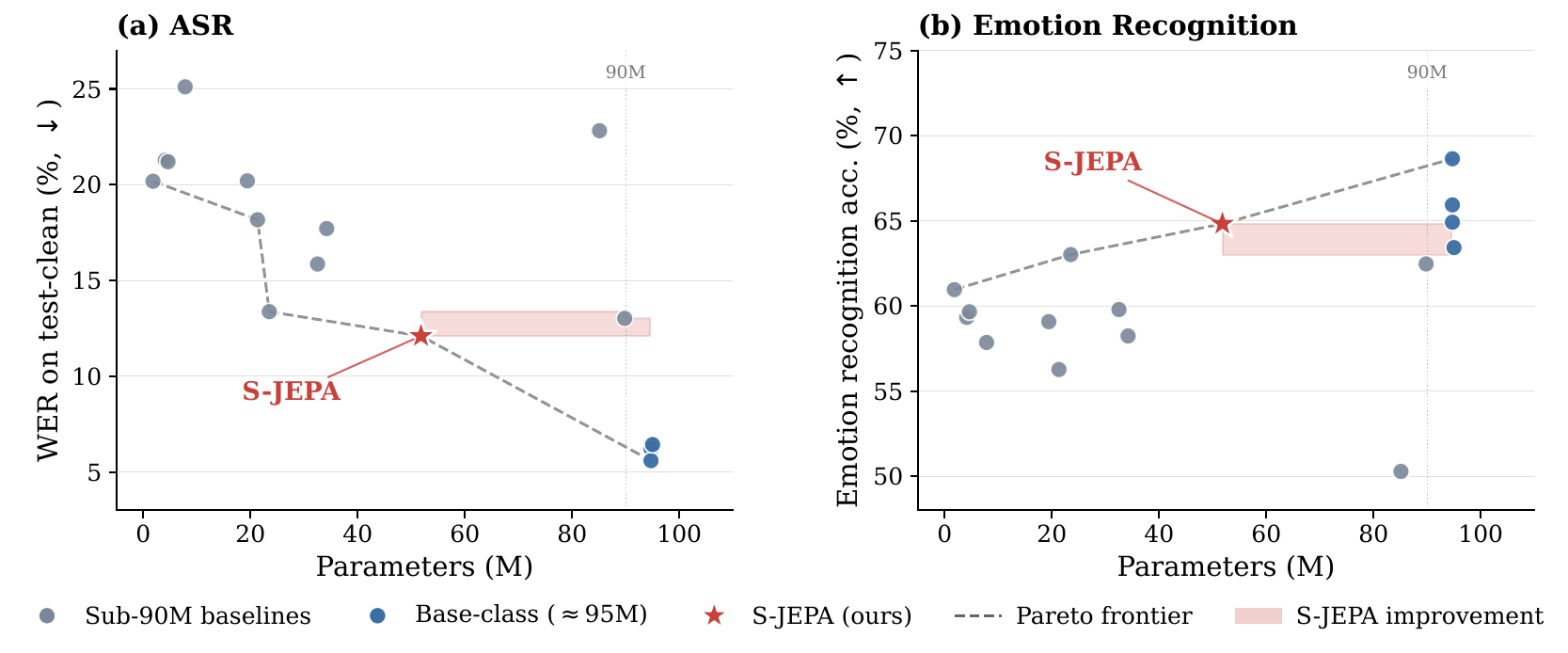}
\caption{\textbf{S-JEPA advances the Pareto frontier on SUPERB below 90M 
parameters.} Parameter count vs.\ performance on (a) ASR (WER on 
LibriSpeech \texttt{test-clean}, $\downarrow$) and (b) emotion 
recognition (accuracy, $\uparrow$). Dashed line: Pareto frontier across 
all evaluated SSL methods. Shaded region: S-JEPA's improvement over the 
prior best in the 51.8M--90M range. S-JEPA reaches 12.10\% WER (vs.\ 
13.02\% for DeCoAR~2.0 at 89.8M) and 64.83\% ER accuracy (matching 
HuBERT-Base at 55\% of its parameters), trained directly without a 
teacher.}
\label{fig:pareto}
\end{figure*}

\begin{table}[t]
\small

\caption{Frozen-encoder evaluation on the SUPERB benchmark: ASR (WER on LibriSpeech \texttt{test-clean}), emotion recognition (ER, accuracy), and slot filling (SF, F1 and CER). All entries use the standard SUPERB protocol with greedy CTC decoding for ASR. Baseline numbers are from the public SUPERB leaderboard. Parameter counts are inference-time encoder only. Best result within the sub-90M block in bold.}
\label{tab:superb}
\centering
\setlength{\tabcolsep}{6pt}
\begin{adjustbox}{width=\columnwidth}
\begin{tabular}{l@{\hspace{1em}}r@{\hspace{1.5em}}c@{\hspace{1em}}c@{\hspace{1em}}c@{\hspace{1em}}c}
\toprule
\multirow{2}{*}{\textbf{Method}} & \multirow{2}{*}{\textbf{\# Params}} & \textbf{ASR} & \textbf{ER} & \textbf{SF} & \textbf{SF} \\
                                 &                                    & WER ($\downarrow$) & Acc.\ ($\uparrow$) & F1 ($\uparrow$) & CER ($\downarrow$) \\
\midrule
\multicolumn{6}{l}{\textit{Sub-90M baselines}} \\
\midrule
modified CPC~\citep{oord2018cpc}              & \phantom{00}1.8M & 20.18 & 60.96 & 71.19 & 49.91 \\
APC~\citep{chung2019apc}                       & \phantom{00}4.1M & 21.28 & 59.33 & 70.46 & 50.89 \\
VQ-APC~\citep{chung2020vqapc}                  & \phantom{00}4.6M & 21.20 & 59.66 & 68.53 & 52.91 \\
PASE+~\citep{pascual2019pase}                  & \phantom{00}7.8M & 25.11 & 57.86 & 62.14 & 60.17 \\
NPC~\citep{liu2020npc}                          & \phantom{0}19.4M & 20.20 & 59.08 & 72.79 & 48.44 \\
TERA~\citep{liu2021tera}                        & \phantom{0}21.3M & 18.17 & 56.27 & 67.50 & 54.17 \\
DistilHuBERT~\citep{chang2022distilhubert}      & \phantom{0}23.5M & 13.37 & 63.02 & 82.57 & 35.59 \\
wav2vec~\citep{schneider2019wav2vec}            & \phantom{0}32.5M & 15.86 & 59.79 & 76.37 & 43.71 \\
vq-wav2vec~\citep{baevski2019vqwav2vec}         & \phantom{0}34.2M & 17.71 & 58.24 & 77.68 & 41.54 \\
Mockingjay~\citep{liu2020mockingjay}            & \phantom{0}85.1M & 22.82 & 50.28 & 61.59 & 58.89 \\
DeCoAR 2.0~\citep{ling2020decoar2}              & \phantom{0}89.8M & 13.02 & 62.47 & 83.28 & 34.73 \\
\textbf{S-JEPA (ours)}                          & \phantom{0}\textbf{51.8M} & \phantom{0}\textbf{12.10} & \textbf{64.83} & \textbf{83.05} & \textbf{33.17} \\
\midrule
\multicolumn{6}{l}{\textit{Base-class ($\approx$95M)}} \\
\midrule
HuBERT Base~\citep{hsu2021hubert}               & \phantom{0}94.7M & \phantom{0}6.42 & 64.92 & 88.53 & 25.20 \\
WavLM Base~\citep{chen2022wavlm}                & \phantom{0}94.7M & \phantom{0}6.21 & 65.94 & 89.38 & 22.86 \\
WavLM Base+~\citep{chen2022wavlm}               & \phantom{0}94.7M & \phantom{0}5.59 & 68.65 & 90.58 & 21.20 \\
wav2vec 2.0 Base~\citep{baevski2020wav2vec}    & \phantom{0}95.0M & \phantom{0}6.43 & 63.43 & 88.30 & 24.77 \\
\midrule
\multicolumn{6}{l}{\textit{Large-class (reference, $\geq$316M)}} \\
\midrule
HuBERT Large~\citep{hsu2021hubert}              & 316.6M           & \phantom{0}3.62 & 67.62 & 89.81 & 21.76 \\
WavLM Large~\citep{chen2022wavlm}               & 316.6M           & \phantom{0}3.44 & 70.62 & 92.21 & 18.36 \\
wav2vec 2.0 Large~\citep{baevski2020wav2vec}   & 316.6M           & \phantom{0}3.75 & 65.64 & 87.11 & 27.31 \\
\bottomrule
\end{tabular}
\end{adjustbox}
\end{table}

\subsection{Frozen-Encoder Results on SUPERB}
\label{sec:results}

Table~\ref{tab:superb} reports frozen-encoder evaluation on three SUPERB tasks: ASR (WER on LibriSpeech \texttt{test-clean}), ER (emotion recognition accuracy), and SF (slot filling F1 and CER). All numbers in the table follow the SUPERB protocol: a small task-specific head is trained on frozen encoder features, and ASR uses greedy CTC decoding. Under this setup, S-JEPA reaches 12.10\% WER and 3.62\% CER. We additionally report S-JEPA's WER under standard LibriSpeech 4-gram LM rescoring ($\alpha = 0.5$, $\beta = 1.5$, beam~$=100$): 8.50\% WER and 2.90\% CER. The greedy number is the headline used for direct comparison with the table baselines.

\paragraph{A new Pareto frontier below 90M parameters.} 
Among the 12 evaluated methods below 90M parameters, S-JEPA dominates on  all three tasks (cf. Figure~\ref{fig:pareto}): lowest WER (12.10\% vs.\ DistilHuBERT's 13.37\% and  DeCoAR~2.0's 13.02\%), highest ER accuracy (64.83\% vs.\ DistilHuBERT's  63.02\%), and SF F1 essentially tied with DeCoAR~2.0 (83.05 vs.\ 83.28) 
at 58\% of its parameter count. Unlike DistilHuBERT, whose 23.5M parameters reflect only the student model and require a 94.7M HuBERT teacher to produce, S-JEPA is trained directly at 51.8M in a single pass. Base-class methods (HuBERT-Base 6.42\%, WavLM-Base 6.21\%, wav2vec 2.0 Base 6.43\%) achieve substantially lower WER at roughly twice S-JEPA's parameter count; we establish a Pareto frontier within the sub-90M regime, not parity with Base-class methods.

\paragraph{Beyond ASR.} S-JEPA reaches 64.83\% accuracy on emotion recognition and 83.05 F1 / 33.17 CER on slot filling. On ER this is essentially tied with HuBERT-Base (64.92\%) at 55\% of HuBERT-Base's parameter count, and exceeds all sub-90M baselines (next-best DistilHuBERT at 63.02\%). On SF, S-JEPA's F1 (83.05) is essentially tied with DeCoAR 2.0 (83.28 at 89.8M) and beats all evaluated methods below DeCoAR 2.0 in parameter count; the strongest Base-class methods (WavLM-Base+ 90.58, WavLM-Base 89.38, HuBERT-Base 88.53) are 5.5--7.5 F1 points ahead, a larger gap than on ASR or ER, consistent with SF being a more demanding task that benefits from the additional capacity of the 95M-parameter encoders.

\subsection{Probing Frozen Representations}
\label{sec:probing}

We further probe S-JEPA representations against HuBERT and WavLM on four LibriSpeech~\citep{panayotov2015librispeech} tasks: Speaker Identification, Gender Classification, Chapter Identification, and Phoneme Classification. Chapter Identification is the most discriminative task for this comparison: chapters within a book are often read by the same speaker, so the available cue is subtle prosodic and recording variation rather than speaker identity alone. Probing protocol details (probe architectures, training, splits) are in Section~\ref{app:probing}.

S-JEPA outperforms WavLM and HuBERT on Speaker ID, Gender, and Chapter ID under linear probes (Table~\ref{tab:probing}); the Speaker ID and Gender margins are at the ceiling of the metric, while the Chapter ID gap (S-JEPA 93.2 vs.\ HuBERT 88.2) is more substantial. On Phoneme classification the linear probe favors WavLM (86.1 vs.\ 82.8 for S-JEPA), but switching to an MLP probe reverses the ranking (S-JEPA 88.0 vs.\ WavLM 87.5). The MLP gain over the linear probe is substantially larger for S-JEPA ($+5.2$ points) than for WavLM ($+1.4$) or HuBERT ($+1.7$), suggesting that S-JEPA encodes phonetic information in a form that is more recoverable nonlinearly.

 \begin{table}[h]
\centering
\caption{\textbf{S-JEPA leads on Speaker, Gender, and Chapter ID, and benefits more from a nonlinear probe on phoneme classification.} Probing accuracy on LibriSpeech~\citep{panayotov2015librispeech} across four tasks. Mean $\pm$ standard deviation across 3 splits. Chapter ID is the most discriminative comparison: chapters within a book are typically read by the same speaker, so the cue is subtle prosodic and recording variation rather than speaker identity.}
\label{tab:probing}
\small
\setlength{\tabcolsep}{8pt}
\renewcommand{\arraystretch}{1.15}
\begin{adjustbox}{width=\columnwidth}
\begin{tabular}{lccc}
\toprule
\textbf{Task (Accuracy $\uparrow$)} & \textbf{WavLM} & \textbf{HuBERT} & \textbf{S-JEPA} \\
\midrule
Speaker ID            & 91.1 $\pm$ 1.0          & 99.1 $\pm$ 0.3          & \textbf{99.7 $\pm$ 0.3} \\
Gender Classification & 96.3 $\pm$ 0.5          & 99.5 $\pm$ 0.3          & \textbf{99.6 $\pm$ 0.3} \\
Chapter ID            & 59.5 $\pm$ 1.0          & 88.2 $\pm$ 1.0          & \textbf{93.2 $\pm$ 1.0} \\
\midrule
Phoneme (linear)      & \textbf{86.1 $\pm$ 0.2} & 84.8 $\pm$ 0.1          & 82.8 $\pm$ 0.1          \\
Phoneme (MLP)         & 87.5 $\pm$ 0.1          & 86.5 $\pm$ 0.1          & \textbf{88.0 $\pm$ 0.1} \\
\bottomrule
\end{tabular}
\end{adjustbox}
\end{table}

\iffalse
\begin{table}[h]
\centering
\caption{Probing accuracy across downstream tasks on LibriSpeech~\citep{panayotov2015librispeech}.}
\label{tab:results}
\small
\begin{tabular}{lccc}
\toprule
\textbf{Task} & \textbf{WavLM} & \textbf{HuBERT} & \textbf{S-JEPA} \\
\midrule
Speaker ID                        & 91.1 $\pm$ 1.0 & 99.1 $\pm$ 0.3 & \textbf{99.7 $\pm$ 0.3} \\
Gender Classification             & 96.3 $\pm$ 0.5 & 99.5 $\pm$ 0.3 & \textbf{99.6 $\pm$ 0.3} \\
Chapter ID                        & 59.5 $\pm$ 1.0 & 88.2 $\pm$ 1.0 & \textbf{93.2 $\pm$ 1.0} \\
Phoneme Linear (3 splits)         & \textbf{86.1 $\pm$ 0.2} & 84.8 $\pm$ 0.1 & 82.8 $\pm$ 0.1 \\
Phoneme MLP (3 splits)            & 87.5 $\pm$ 0.1 & 86.5 $\pm$ 0.1 & \textbf{88.0 $\pm$ 0.1} \\
\bottomrule
\end{tabular}
\end{table}
\fi

%==============================================================================
\vspace{-0.2em}
\begin{figure*}[!t]
\centering
\begin{subfigure}[t]{0.55\linewidth}
    \centering
    \includegraphics[width=\linewidth]{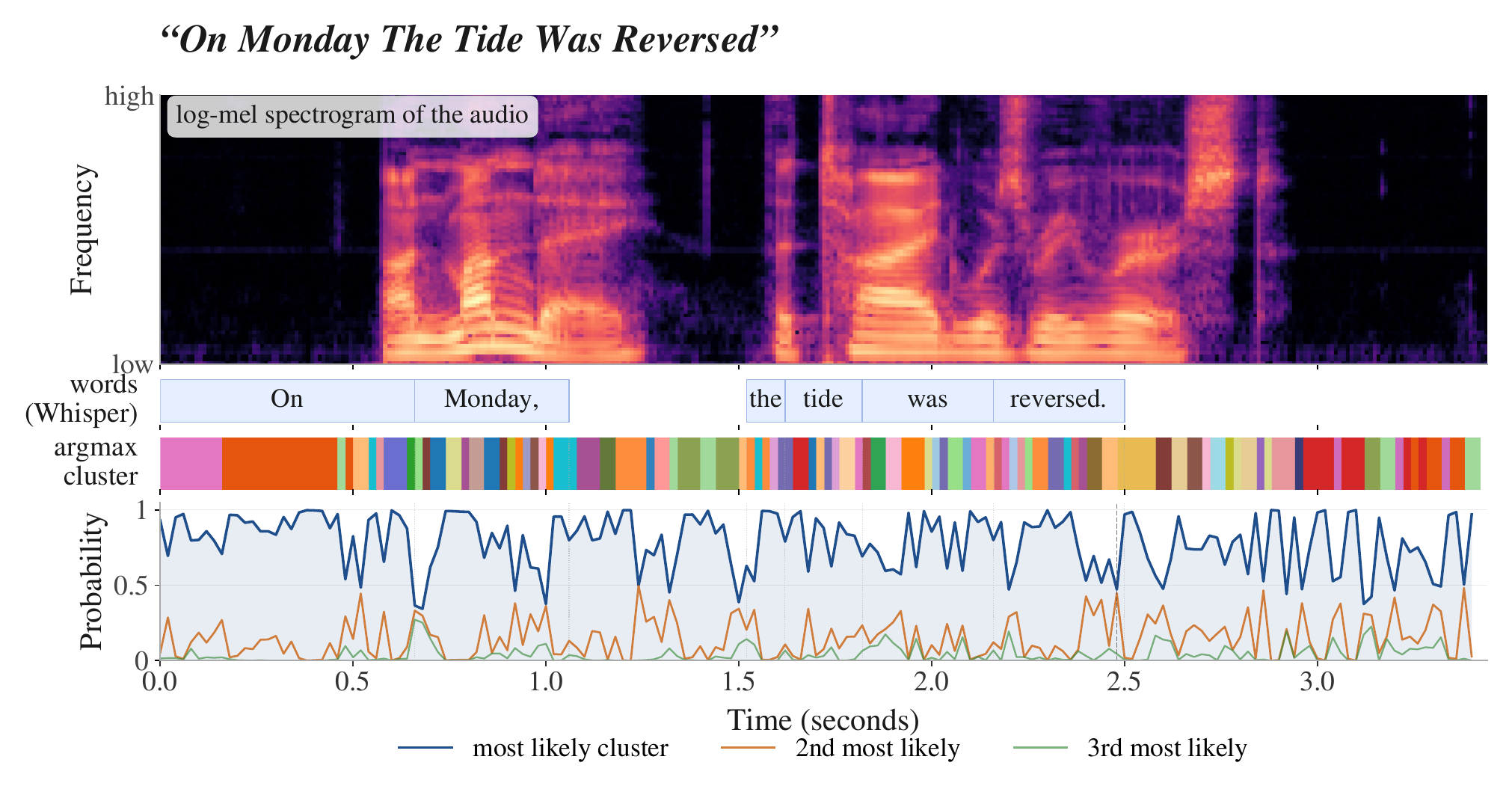}
    \caption{One held-out utterance. From top: log-mel spectrogram; word-level Whisper alignments; argmax cluster ID per $20$\,ms frame ($K\!=\!500$); the predictor's top-three cluster posteriors.}
    \label{fig:utterance}
\end{subfigure}\hfill
\begin{subfigure}[t]{0.42\linewidth}
    \centering
    \includegraphics[width=\linewidth]{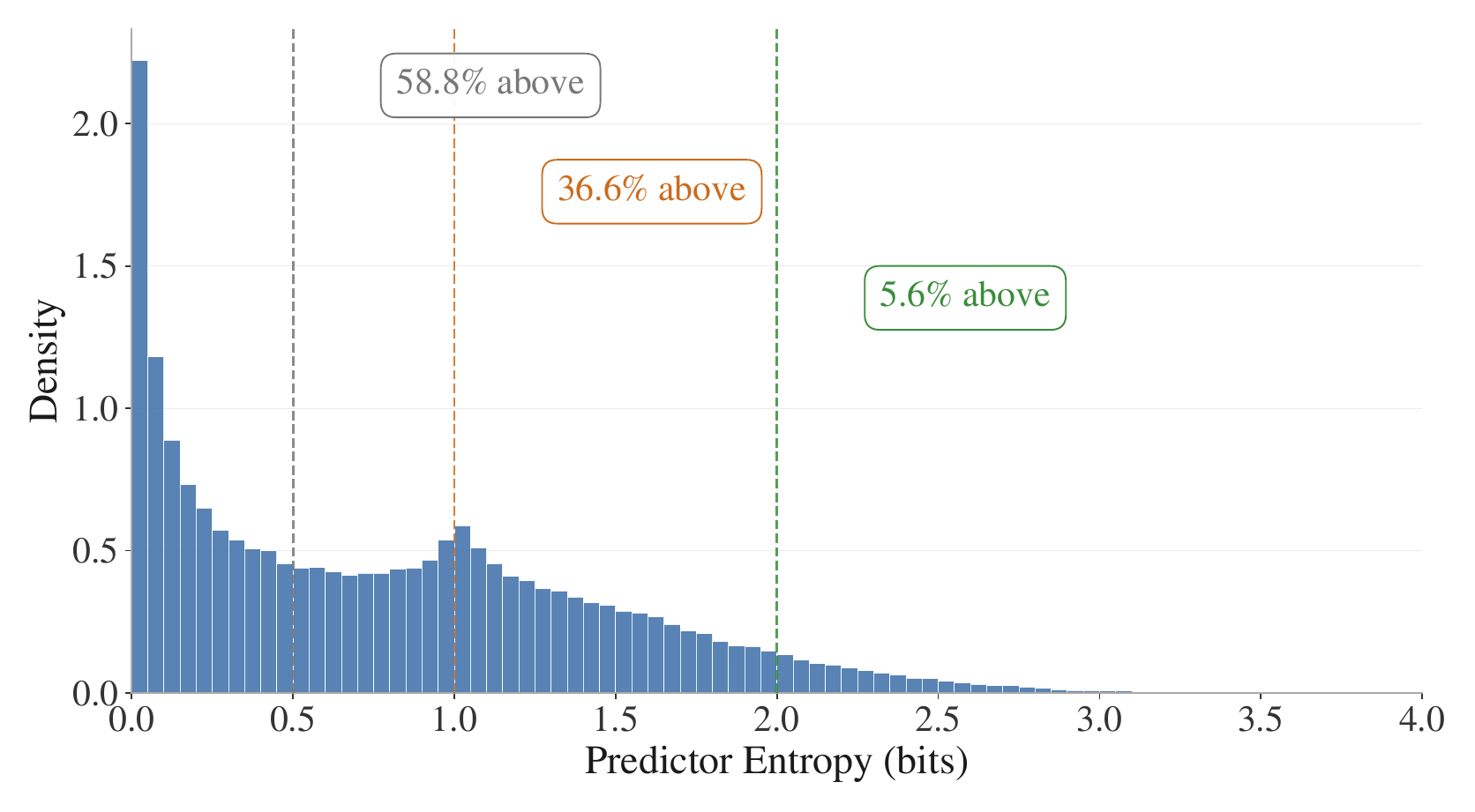}
    \caption{Per-frame predictor entropy across $\sim$153K frames ($N\!=\!500$ \texttt{dev-clean} utterances). Reference verticals at $0.5$, $1$, and $2$\,bits.}
    \label{fig:entropy_distribution}
\end{subfigure}
\caption{\textbf{The soft-target objective produces structured boundary uncertainty, not diffuse fuzziness.} \subref{fig:utterance} On a single utterance, rank-1 posteriors sit near $1.0$ within phonetically stable regions and drop to the $0.3$--$0.45$ range near word boundaries with rank-2 rising to compete. \subref{fig:entropy_distribution} The same pattern at the population level: a bimodal entropy distribution with a confident regime below $0.3$\,bits and a clear secondary mode at $\sim1$bit (the entropy of a perfect two-way tie); $36.6\%$ of frames carry more than $1$\,bit.}
\label{fig:soft_target_behavior}
\end{figure*}

\section{Analysis}
\label{sec:analysis}
%==============================================================================

We analyse the learned representations through two lenses: a per-utterance walk-through that makes the soft-target behaviour visible at the frame level, and an aggregate measurement that quantifies it.

%------------------------------------------------------------------------------
\subsection{Per-utterance Interpretability}
\label{sec:per_utterance}
%------------------------------------------------------------------------------

We extract per-frame posteriors from the frozen encoder and cluster head on a single held-out utterance from LibriSpeech \texttt{dev-clean}, with masking disabled. At each 20\,ms frame $t$ we obtain the predictor's distribution $p_t$ over the $K = 500$ Phase-2 GMM components. To label the time axis we run an off-the-shelf small Whisper~\citep{radford2022robustspeechrecognitionlargescale} model for word-level start/end times. Whisper alignments are used only as a visual reference and play no role in S-JEPA training or in any quantitative results reported in the paper. The argmax strip in Figure~\ref{fig:utterance} also shows that with $K = 500$ clusters at 20\,ms resolution, cluster IDs change at sub-phonemic rates: the encoder is learning a finer-than-phonetic acoustic vocabulary, consistent with our choice to anchor on encoder features rather than phoneme-aligned labels.

\subsection{Aggregate Distribution of Predictor Uncertainty}
\label{sec:aggregate}
%------------------------------------------------------------------------------

To test whether the soft-target behaviour generalises beyond a single utterance, we sample $N = 500$ utterances from \texttt{dev-clean} \cite{panayotov2015librispeech}, extract per-frame predictor distributions $\{p_t\}$ with masking disabled, and convert each frame's entropy to bits, $H_t = -\sum_{k=1}^{K} p_{t,k} \log_2 p_{t,k} \in [0, \log_2 K]$. Reporting in bits gives reference values with fixed semantic meaning that do not depend on $K$: $H = 1$ bit corresponds to a perfect two-way 50/50 tie, $H = 2$ bits to a four-way uniform spread, and $H = \log_2 K = 9$ bits to a uniform distribution over all $K = 500$ clusters.

The bimodality is the load-bearing observation. A model whose softness is just diffuse fuzziness would produce a unimodal distribution sliding from low to high entropy. A model whose softness was a degenerate fallback to uniform priors would pile mass near $\log_2 K$. S-JEPA does neither: it produces a confident posterior on most frames and a structured two-way competition on a substantial minority. Specifically (cf. Figure~\ref{fig:entropy_distribution}), the per-frame entropy is bimodal, with a confident regime concentrated below 0.3~bits and a clear secondary mode near 1~bit corresponding to two-way ties between competing clusters; 36.6\% of frames carry more than 1~bit of entropy. Over a third of frames are in a regime where a hard-target objective would have to commit to one cluster ID and discard the inter-cluster mass.

%==============================================================================
\section{Discussion}
%==============================================================================

S-JEPA can be read as soft-target HuBERT in a JEPA-style architecture: GMM posteriors replace $k$-means hard labels, and in doing so, they preserve the acoustic ambiguity at the category boundaries~\citep{ladefoged2014course,stevens1998acoustic} that hard assignments collapse, and the predictor-fills-the-mask pattern from JEPA work~\citep{lecun2022jepa,assran2023ijepa,fang2024ajepa} sits on top of a categorical prediction target. Beyond eliminating the stop-restart pattern of HuBERT and WavLM, the online + adaptive-layer design also removes a discrete tuning decision present in the offline-clustering setup: the choice of which intermediate transformer layer to cluster on no longer has to be made by hand and held fixed for an entire training stage.

The paper's two strongest empirical findings sit alongside each other. S-JEPA is the lowest-WER SSL method below 90M parameters and matches HuBERT-Base on emotion recognition at 55\% of its parameter count (cf. Section~\ref{sec:results}); separately, 36.6\% of the predictor's per-frame posteriors are bimodally structured around a two-way-tie regime that a hard-target objective could not represent (cf. Section~\ref{sec:aggregate}). The hypothesis that soft categorical targets carry information that argmax labels discard is the canonical argument of knowledge distillation~\citep{hinton2015distilling} and recurs in self-distillation with softened teacher distributions over discrete prototypes~\citep{caron2021dino}. S-JEPA's targets are also discrete --- soft posteriors over $K{=}500$ GMM components --- and the bimodality finding suggests the inter-cluster mass distinguishing a soft from a hard target is the information-carrying part of the signal on those frames. A controlled head-to-head ablation against an equivalent hard-target baseline at matched compute is a natural next step we leave to future work.

\textbf{Limitations and Future Work}
\label{sec:limitations}
The periodically switched EMA schedule lacks a controlled comparison against a more principled approach. 
All training and evaluation are in English; extension to tonal languages, low-resource regimes, and non-speech audio is open. GMM is promising as an alternative to SIGReg in JEPA formulations~\citep{balestriero2025lejepa}, and is powerful because it can learn specialized regions in the latent space~\citep{goldfeder2026ai} without enforcing a single gaussian distribution, but the exact relationship should be explored in further work.

%==============================================================================
\section{Conclusion}
%==============================================================================
%v1

We introduce S-JEPA, a soft-target masked-prediction objective for self-supervised speech representation learning: GMM posteriors at masked positions matched via KL divergence, with the two-phase recipe running as a single continuous training pass via online GMM updates, adaptive layer selection, and a periodically switched EMA decay. With a 51.8M-parameter encoder, S-JEPA is the lowest-WER SSL method below 90M parameters on LibriSpeech \texttt{test-clean} and matches HuBERT-Base on emotion recognition at 55\% of its parameter count. Our analysis of held-out predictor entropy reveals a bimodally structured distribution, with a substantial minority of frames in a two-way-competition regime where a hard-target objective could not represent the inter-cluster ambiguity.

\bibliographystyle{plainnat}
\bibliography{references}

%%%%%%%%%%%%%%%%%%%%%%%%%%%%%%%%%%%%%%%%%%%%%%%%%%%%%%%%%%%%

\newpage
\onecolumn
\appendix

\section{Architecture Details}
\label{app:arch}

The architecture follows HuBERT~\citep{hsu2021hubert} for the encoder, with a single change: 6 Transformer~\citep{vaswani2017attention} encoder layers in place of HuBERT-Base's 12. The predictor and cluster head are described below; both are used only during training and discarded afterwards.

\subsection{CNN Feature Extractor}

A 7-layer 1D convolutional feature extractor maps the raw 16\,kHz waveform to a sequence of frame features. Kernel sizes and strides match HuBERT exactly: $(10,5)$ for the first layer, $(3,2)$ for layers 2--5, and $(2,2)$ for layers 6--7. Each conv is followed by GELU; the first layer additionally uses fp32 group normalization. The total stride is 320 samples, yielding 20\,ms frames at 16\,kHz. Output channel width is 512. Following HuBERT, gradients flowing back through the CNN feature extractor are scaled by $\gamma_{\text{feat}}\!=\!0.1$ to prevent the frontend from learning too quickly.

\subsection{Transformer Stack}

Six Transformer encoder layers, configured following HuBERT-Base except for layer count: 768-dimensional hidden state, 12 attention heads (head dimension 64), feed-forward dimension 3072, GELU activations, post-norm layout. Layer normalization is applied in fp32. A convolutional positional encoding (kernel 128, 16 groups, weight-normalized) is added to the encoder input before the Transformer stack runs. LayerDrop is applied at probability $0.05$ per layer during training. Attention dropout, residual dropout, and activation dropout are each set to $0.1$.

\subsection{Encoder Mask Handling}

The encoder accepts a binary frame-level mask alongside the waveform. After the CNN frontend and the post-extract projection, masked frames are zeroed out before the convolutional positional encoding and the Transformer stack run. The encoder thus produces representations of the visible context with no signal at masked positions; the mask token is injected later, by the predictor.

\subsection{Predictor and Cluster Head}

The predictor injects a learned 768-dimensional mask token at masked positions, adds learned positional embeddings (capacity for 750 frames, sufficient for 15\,s of audio at 20\,ms frames), and processes the result with a single Transformer encoder layer (8 heads, feed-forward dimension $2 \times 768 = 1536$, dropout $0.1$, batch-first). A linear projection produces frame-level features that are passed to the cluster head.

The cluster head is a 3-layer MLP applied to per-frame features:
\begin{equation*}
\mathrm{Linear}(768 \to 768) \to \mathrm{GELU} \to \mathrm{Linear}(768 \to 768) \to \mathrm{GELU} \to \mathrm{Linear}(768 \to K),
\end{equation*}
where $K\!=\!100$ in Phase 1 and $K\!=\!500$ in Phase 2. The cluster head is shared between the visible-position path (encoder output, computed for analysis but unsupervised) and the masked-position path (predictor output, supervised by the KL loss in \Cref{eq:loss}).

\subsection{EMA Encoder}

An architecturally identical copy of the encoder, $\bar{f}$, is maintained throughout training with parameters updated by exponential moving average of the online encoder, in line with self-distillation practice~\citep{grill2020byol, baevski2022data2vec}. In Phase 1 it is unused; in Phase 2 its outputs at the active layer $\ell^\star$ provide the input features on which the online GMM is fit, and its decay rate is switched periodically between two values as in \Cref{eq:ema_switch}.

\subsection{Masking}

Block-wise temporal masking samples contiguous spans until a target fraction of positions is masked. The encoder zeros out masked positions internally; the learned mask token is applied later, by the predictor, before its Transformer layers run.

\section{Training algorithm}
Algorithm \ref{alg:training} summarizes the full procedure.

\begin{algorithm}[t]
\small
\caption{S-JEPA training. Phase~1 uses a fixed MFCC GMM; Phase~2 uses the online encoder-feature GMM with adaptive layer selection and switched EMA decay.}
\label{alg:training}
\begin{algorithmic}[1]
\STATE \textbf{Init:} $f_\phi$, $h_\psi$ (with mask token + pos.\ embeddings), $g_\omega$, EMA encoder $\bar{f}$, GMM, active layer $\ell^\star$, EMA decay $\alpha_t \gets \alpha_{\text{fast}}$
\FOR{each minibatch $\{\mathbf{x}\}$}
    \STATE $\mathbf{x}_{\text{aug}}, \mathbf{x}_{\text{clean}} \gets \text{Augment}(\mathbf{x})$
    \STATE $\mathbf{m} \gets$ MFCC$(\mathbf{x}_{\text{clean}})$ (Phase 1) \textbf{or} $\bar{f}(\mathbf{x}_{\text{clean}})$ at $\ell^\star$ (Phase 2) \COMMENT{no\_grad}
    \STATE $\mathbf{q} \gets \text{GMM.posterior}(\mathbf{m})$
    \IF{Phase 2}
        \STATE GMM $\gets$ EMA-update from $\mathbf{m}$, $\mathbf{q}$ \COMMENT{online; no gradient}
        \IF{layer-check step}
            \STATE Compute $\mathrm{erank}(Z^{(\ell)})$ for all $\ell$; EMA-update per-layer scores; $\ell^\star \gets \arg\max_\ell \mathrm{erank}^{\text{ema}}_\ell$
        \ENDIF
    \ENDIF
    \STATE $\mathit{mask} \gets \text{SampleBlockMask}()$;\quad $z \gets f_\phi(\mathbf{x}_{\text{aug}}, \mathit{mask})$;\quad $\hat{z} \gets h_\psi(z, \mathit{mask})$
    \STATE $\mathbf{p}_t \gets \text{softmax}(g_\omega(\hat{z}_t))$ for $t \in \mathcal{M}$
    \STATE $\mathcal{L} \gets \tfrac{1}{|\mathcal{M}|} \sum_{t \in \mathcal{M}} \text{KL}(\mathbf{q}_t \| \mathbf{p}_t)$; update $\phi, \psi, \omega$ via $\nabla \mathcal{L}$
    \IF{Phase 2}
        \STATE Flip $\alpha_t \in \{\alpha_{\text{fast}}, \alpha_{\text{slow}}\}$ per switching schedule (\Cref{eq:ema_switch});\, $\bar{\phi} \gets \alpha_t \bar{\phi} + (1-\alpha_t) \phi$
    \ENDIF
\ENDFOR
\STATE \textbf{Return:} $f_\phi$ \COMMENT{predictor, cluster head, GMM discarded}
\end{algorithmic}
\end{algorithm}
\section{Phase 2 Transition: Masked-Only Loss and Augmentation Off}
\label{app:phase2_transition}

We started Phase 2 with the same loss configuration as Phase 1: KL applied at both masked and visible positions, with denoising augmentation enabled. Partway through Phase 2 we made two changes simultaneously, motivated by tracking downstream WER on a held-out development set:

\begin{enumerate}
    \item \textbf{Dropping the visible-position loss.} Visible-position cluster predictions are an easier prediction problem (the encoder sees the input directly there) and may supply gradients that are largely redundant with what the masked-position loss already provides. Dropping the visible-position term ($\mathcal{S} = \mathcal{M}$) is consistent with the masked-only configuration used by data2vec~\citep{baevski2022data2vec}, wav2vec~2.0~\citep{baevski2020wav2vec}, and HuBERT~\citep{hsu2021hubert}.
    \item \textbf{Disabling augmentation.} We set $p_{\text{noise}} = p_{\text{mix}} = 0$, so the encoder sees the same clean waveform that the EMA encoder feeds to the GMM. This is in tension with WavLM's~\citep{chen2022wavlm} finding that denoising augmentation is helpful, and likely reflects an interaction specific to our setup: if the GMM target is computed from clean audio while the encoder sees augmented audio, there is an extra invariance demand on the encoder that may make masked-position prediction harder than necessary at this stage of training.
\end{enumerate}

After both changes, downstream WER continued to decrease and the auto-layer mechanism settled at $\ell^\star\!=\!4$. We did not run a controlled ablation toggling each change independently; see \Cref{sec:limitations}.

\section{Training Hyperparameters}
\label{app:hyperparams}

\Cref{tab:hyperparams} reports the full set of training hyperparameters. Values are taken from the launch commands actually used to produce the checkpoints reported in this paper.

\begin{table}[h]
\caption{Training hyperparameters used in our experiments. AdamW~\citep{loshchilov2019adamw} is the optimizer for both phases.}
\label{tab:hyperparams}
\centering
\small
\begin{adjustbox}{width=\textwidth}
\begin{tabular}{l@{\hspace{1.5em}}l@{\hspace{1.5em}}l}
\toprule
& \textbf{Phase 1 (MFCC GMM)} & \textbf{Phase 2 (online GMM)} \\
\midrule
\multicolumn{3}{l}{\emph{Encoder backbone (shared across phases)}} \\
Hidden dimension                          & 768  & 768  \\
Transformer layers                        & 6    & 6    \\
Attention heads                           & 12   & 12   \\
FFN dimension                             & 3072 & 3072 \\
CNN feature-extractor channels            & 512  & 512  \\
CNN total stride / frame rate             & 320 / 50\,Hz & 320 / 50\,Hz \\
Conv positional encoding (kernel, groups) & (128, 16) & (128, 16) \\
Dropout / attention / activation          & 0.1 / 0.1 / 0.1 & 0.1 / 0.1 / 0.1 \\
LayerDrop                                 & 0.05 & 0.05 \\
Feature-extractor gradient multiplier     & 0.1  & 0.1  \\
\midrule
\multicolumn{3}{l}{\emph{GMM}} \\
$K$ (number of components)                & 100   & 500   \\
Feature space                             & 39-dim MFCC+$\Delta$+$\Delta\Delta$ & 768-dim EMA encoder features \\
Fitting                                   & k-means init (5 it.) + EM (20 it.) & online EMA of suff.\ stats \\
Online param EMA decay                    & --- (frozen) & 0.999 \\
Layer-check cadence (auto-layer)          & --- (n/a)    & every $\sim 10{,}000$ steps \\
\midrule
\multicolumn{3}{l}{\emph{Optimization}} \\
Optimizer                                 & AdamW & AdamW \\
$\beta_1, \beta_2$                        & $(0.9, 0.99)$ & $(0.9, 0.99)$ \\
Weight decay                              & $10^{-3}$ & $10^{-3}$ \\
Learning rate                             & $1\!\times\!10^{-4}$ & $2.5\!\times\!10^{-5}$ \\
Batch size (global)                       & 192 & 192 \\
GPUs                                      & 8 & 8 \\
Max input duration                        & 15\,s & 15\,s \\
\midrule
\multicolumn{3}{l}{\emph{EMA encoder $\bar{f}$ (periodically switched decay, Phase 2 only)}} \\
$\alpha_{\text{slow}}$                    & --- (n/a) & 0.9999 \\
$\alpha_{\text{fast}}$                    & --- (n/a) & 0.999 \\
Switching cadence                         & --- (n/a) & every $\sim 20{,}000$ steps (avg.) \\
\midrule
\multicolumn{3}{l}{\emph{Masking}} \\
Mask ratio                                & 0.65 & 0.65 \\
Mask span length (frames)                 & 10   & 10   \\
\midrule
\multicolumn{3}{l}{\emph{Loss configuration}} \\
KL applied at                             & masked + visible & masked + visible (early), masked only (after) \\
\midrule
\multicolumn{3}{l}{\emph{Augmentation}} \\
$p_{\text{noise}}$                        & 0.25 & 0.25 (early), 0.0 (after) \\
$p_{\text{mix}}$                          & 0.25 & 0.25 (early), 0.0 (after) \\
SNR range (noise mix), dB                 & $(-5, 20)$ & $(-5, 20)$ when active \\
Energy ratio range (utterance mix), dB    & $(-5, 5)$  & $(-5, 5)$ when active \\
\bottomrule
\end{tabular}
\end{adjustbox}
\end{table}

\section{GMM Details}
\label{app:gmm}

\subsection{Posterior Computation}

For the diagonal-covariance GMM the posterior is computed in closed form~\citep{bishop2006pattern}:
\begin{equation}
q_k(\mathbf{m}) = \text{softmax}_k\left(\log \pi_k - \frac{D}{2}\log(2\pi) - \frac{1}{2}\sum_{d=1}^{D}\left[\log \sigma_{k,d}^2 + \frac{(m_d - \mu_{k,d})^2}{\sigma_{k,d}^2}\right]\right)
\end{equation}
Variances are clamped at a small floor ($10^{-6}$) for numerical stability throughout.

\subsection{Phase 1 GMM Fitting}

The Phase 1 GMM is fit once on 39-dimensional MFCC features (13 cepstral coefficients with first and second deltas, 20\,ms frames) and held fixed thereafter. Fitting proceeds in three steps.

\textbf{Feature collection via reservoir sampling.} We stream audio from the training corpus and extract MFCC+delta features frame by frame. To bound memory, we maintain a frame reservoir: incoming frames are added until the reservoir reaches a target size, after which each new frame replaces a uniformly random reservoir entry with the appropriate probability~\citep{vitter1985reservoir}.

\textbf{Mini-batch k-means initialization.} The GMM means are initialized by running mini-batch k-means~\citep{macqueen1967kmeans} over the reservoir for 5 iterations. Cluster variances are initialized as the per-cluster diagonal sample variances of their k-means members, and mixture weights as the empirical cluster proportions. Empty clusters are re-seeded from random reservoir frames.

\textbf{EM refinement.} Starting from the k-means initialization, we run 20 iterations of expectation-maximization~\citep{dempster1977em} for the full diagonal-covariance GMM. Sufficient statistics are accumulated in chunks over the reservoir to avoid materializing the full $N \!\times\! K$ responsibility matrix; $M$-step updates are applied at the end of each iteration.

\subsection{Online GMM Updates (Phase 2)}

In Phase 2, the GMM parameters $\{\pi_k, \boldsymbol{\mu}_k, \boldsymbol{\sigma}_k^2\}$ are updated online from minibatch sufficient statistics. Given $N$ feature vectors $X \in \mathbb{R}^{N \times D}$ from the EMA encoder's active layer, with their soft posteriors $Q \in \mathbb{R}^{N \times K}$, we compute responsibility-weighted means $\hat{\boldsymbol{\mu}}_k = \frac{1}{\hat{N}_k}\sum_n q_{n,k} \mathbf{x}_n$, variances $\hat{\boldsymbol{\sigma}}_k^2 = \frac{1}{\hat{N}_k}\sum_n q_{n,k} (\mathbf{x}_n - \hat{\boldsymbol{\mu}}_k)^2$, and weights $\hat{\pi}_k = \hat{N}_k / \sum_j \hat{N}_j$ where $\hat{N}_k = \sum_n q_{n,k}$. We then EMA-update:
\begin{equation}
\theta \leftarrow \alpha \theta + (1-\alpha) \hat{\theta}, \quad \theta \in \{\pi, \boldsymbol{\mu}, \boldsymbol{\sigma}^2\}
\end{equation}
No gradient flows through the GMM at any point.

\subsection{Chunked Soft Assignment}

For large $K$ and large batch sizes, soft posteriors are computed in chunks over $K$ to bound peak memory:

\begin{algorithm}[h]
\caption{Chunked Soft Assignment}
\begin{algorithmic}[1]
\STATE \textbf{Input:} Features $X \in \mathbb{R}^{N \times D}$, GMM parameters $\{\pi_k, \mu_k, \sigma_k^2\}$, chunk sizes
\STATE \textbf{Output:} Posteriors $Q \in \mathbb{R}^{N \times K}$
\FOR{$n = 0$ to $N$ step $\text{batch\_size}$}
    \FOR{$k = 0$ to $K$ step $\text{chunk\_k}$}
        \STATE $\Delta \gets X[n:n'] - \mu[k:k']$
        \STATE $\text{mahal} \gets \sum_d \Delta_d^2 / \sigma_{k,d}^2$
        \STATE $\log p_{n,k} \gets -\frac{D}{2}\log(2\pi) - \frac{1}{2}\sum_d \log \sigma_{k,d}^2 - \frac{1}{2}\text{mahal}$
    \ENDFOR
    \STATE $Q[n:n'] \gets \text{softmax}(\log \pi + \log p)$
\ENDFOR
\end{algorithmic}
\end{algorithm}

\section{Periodically Switched EMA Decay}
\label{app:cyclic-ema}

This appendix gives the full details of the periodically switched EMA decay on the EMA encoder $\bar{f}$ used in Phase 2 (see \Cref{par:cyclic-ema} for the conceptual description).

\paragraph{Schedule.} The decay rate $\alpha_t$ is a piecewise-constant function of the step index taking one of two values. We hold the rate fixed at one value for some number of steps, then flip it to the other, then back. Intervals are not exactly uniform across training, but they average $T_{\text{switch}} \approx 20{,}000$ steps per interval. The target-encoder update is $\bar{\phi} \leftarrow \alpha_t \bar{\phi} + (1-\alpha_t)\phi$, applied once per minibatch. We use $\alpha_{\text{fast}} = 0.999$ and $\alpha_{\text{slow}} = 0.9999$. There is no smoothing or interpolation between values.

\paragraph{Effective EMA timescale.} The characteristic timescale at decay $\alpha$ is $\tau(\alpha) = 1/(1-\alpha)$ steps. With our values, $\tau(\alpha_{\text{slow}}) = 10{,}000$ and $\tau(\alpha_{\text{fast}}) = 1{,}000$. The mean switching cadence ($\sim 20{,}000$ steps) is comfortably larger than either timescale, so within a single fast-rate interval the EMA encoder has time to substantially track the online encoder, and within a single slow-rate interval the EMA encoder has time to settle and provide approximately stationary targets.

\paragraph{Interaction with online GMM updates.} The online GMM is EMA-updated at every step from the EMA encoder's outputs at the active layer. When the schedule is at $\alpha_{\text{fast}}$, the EMA encoder shifts faster, so the responsibility-weighted statistics that drive the GMM update reflect a more recent encoder, and the GMM's components track recent encoder improvements. When the schedule is at $\alpha_{\text{slow}}$, the EMA encoder is approximately stationary, and so is the distribution from which the GMM's sufficient statistics are sampled.

\paragraph{Initialization and phase boundary.} At the start of Phase 2, $\bar{f}$ is initialized from the Phase 1 encoder ($\bar{\phi} \leftarrow \phi$). Phase 2 begins under $\alpha_{\text{fast}}$: at the Phase 1$\to$2 boundary the online encoder has just been trained against MFCC GMM targets, and the online GMM (just initialized over EMA encoder features) benefits from being able to track the online encoder closely as it begins adapting to the new $K\!=\!500$ targets.

\section{Effective Rank and Adaptive Layer Selection}
\label{app:erank_calibration}

\paragraph{Computation.} For each transformer layer $\ell$, we sample $N\!=\!2{,}000$ frame embeddings from the EMA encoder's output at layer $\ell$, center them, compute the singular values via SVD, normalize them to a probability distribution, and apply \Cref{eq:erank}. The per-layer values are EMA-smoothed across measurement intervals.

\paragraph{Cadence.} Layer checks are performed at a fixed step cadence during Phase 2, set to roughly the natural EMA decay timescale (i.e., the number of steps over which the EMA-smoothed score is dominated by recent measurements). When the argmax changes, the GMM input layer switches at the start of the next minibatch with no other training discontinuity.

\section{Probing Protocol}
\label{app:probing}
 
This appendix describes the probing protocol used in Section~\ref{sec:probing}. All probes are trained on frozen encoder features extracted from each method (S-JEPA, HuBERT-Base, WavLM-Base), with no fine-tuning of the encoder.
 
\paragraph{Features.} For each utterance, 768 dimensional frame-level features are extracted from the final encoder layer at 20\,ms resolution. For utterance-level tasks (Speaker ID, Gender, Chapter ID), frame features are mean-pooled over the utterance. For Phoneme Classification, frame features are used directly without pooling, with each frame assigned the phoneme label of the interval containing its center timestamp, using forced alignments from the \texttt{gilkeyio/librispeech-alignments} dataset.

\paragraph{Linear probe.} We use $L_2$-regularized multinomial logistic regression (scikit-learn's \texttt{LogisticRegression}, $C=1.0$, lbfgs solver, up to 2000 iterations). Features are standardized to zero mean and unit variance using training-split statistics. We use the same probe configuration across all three encoders (S-JEPA, HuBERT-Base, WavLM-Base).
 
\paragraph{MLP probe.} We used an MLP with 2 hidden layers each with 256 dimensions, ReLU activations, and 0.1 dropout after each hidden layer. We used ADAM, lr = 1e-3, and CrossEntropyLoss with a batchsize of 512, for 20 epochs.

\paragraph{Splits.} All probes are trained and evaluated on LibriSpeech \texttt{dev-clean} using random 80/20 train--test splits. Utterance-level tasks (Speaker ID, Gender, Chapter ID) use ten splits; frame-level Phoneme Classification uses three. Reported numbers in Table~\ref{tab:probing} are mean $\pm$ standard deviation across splits.

\paragraph{Tasks.} Speaker ID is multiclass classification over the speakers in LibriSpeech \texttt{dev-clean}. Gender Classification is binary (male/female). Chapter ID is multiclass classification over book chapters; since each chapter is read by a single speaker but speakers read multiple chapters, distinguishing chapters by the same speaker requires recording-session and prosodic cues beyond speaker identity. Phoneme Classification is frame-level multiclass classification over the 67 phonemes in the forced alignments.

\section{Augmentation Details}
\label{app:augment}

The denoising augmentation scheme follows WavLM~\citep{chen2022wavlm}. The encoder sees the augmented waveform; the GMM-target side computes its features from the clean waveform.

\subsection{Energy-Based SNR Mixing}

Given clean signal $x$ and noise $n$, mixing at target SNR:
\begin{align}
E_x &= \frac{1}{|x|}\sum_i x_i^2, \quad E_n = \frac{1}{|n|}\sum_i n_i^2, \\
\alpha &= \sqrt{\frac{E_x}{10^{\text{SNR}/10} \cdot E_n}}, \\
x_{\text{aug}} &= x + \alpha \cdot n
\end{align}

\subsection{Utterance Mixing}

A segment from another utterance is mixed with energy-based weighting:
\begin{equation}
\mathbf{x}_{\text{mix}}[t_1:t_2] = \mathbf{x}_1[t_1:t_2] + \beta \cdot \mathbf{x}_2[s_1:s_2],
\end{equation}
where $\beta = \sqrt{E_1 \cdot 10^{\rho/10} / E_2}$ with energy ratio $\rho$ sampled from a fixed range. Concretely:
\begin{enumerate}
    \item Sample mix length $L_{\text{mix}}$ as a random fraction of the primary length.
    \item Sample start positions $t_1, t_2$ in the primary and secondary signals.
    \item Extract regions $r_1 \gets x_1[t_1:t_1+L]$, $r_2 \gets x_2[t_2:t_2+L]$ and compute energies $E_1, E_2$.
    \item Sample energy ratio $\rho \sim \text{Uniform}$ over a fixed dB range.
    \item Compute $\beta$ and mix $x_1[t_1:t_1+L] \gets r_1 + \beta \cdot r_2$.
\end{enumerate}

\section{SUPERB Evaluation Protocol}
\label{app:superb}

We follow the standard SUPERB~\citep{yang2021superbspeechprocessinguniversal} protocol: the pre-trained encoder is frozen and small task-specific heads are trained on top. Across tasks, the encoder receives raw waveform input and the task-specific head consumes its frame-level outputs (or a learned weighted combination across encoder layers, depending on the task). Specific task setups, evaluation metrics, and reported numbers are filled in alongside \Cref{tab:superb}.

\end{document}